\documentstyle[12pt]{article}
\input epsf
\begin{document}
\pagestyle{plain}
\setcounter{page}{1}
\newcounter{bean}
\baselineskip16pt

%--------+---------+---------+---------+---------+---------+---------+

\begin{titlepage}
\begin{flushright}
PUPT-1969\\
hep-th/0011179
\end{flushright}

\vspace{20 mm}

\begin{center}
{\huge A two point function in a cascading ${\cal N}=1$ gauge theory}
\vspace{5mm}
{\huge from supergravity}
\end{center}
\vspace{10 mm}
\begin{center}
{\large  
Michael Krasnitz\\
}
\vspace{3mm}
Joseph Henry Laboratories\\
Princeton University\\
Princeton, New Jersey 08544
\end{center}
\vspace{2cm}
\begin{center}
{\large Abstract}
\end{center}
\noindent

We compute the two point correlation function of a dimension 4 operator in a nonconformal cascading ${\cal N}=1$ SUSY gauge theory using the supergravity dual found by Klebanov and Strassler \cite{ks}. The two point function has a logarithmic correction to conformal behavior which is related to the scale dependence of the effective number of colors. The nonsingular behavior of this correlator suggests that the theory remains a local 4-dimensional quantum field theory at all scales. We also compute the spectrum of low-lying glueball modes corresponding to the above operator.

\vspace{2cm}
\begin{flushleft}
November 2000

\end{flushleft}
\end{titlepage}

%--------+---------+---------+---------+---------+---------+---------+

\newpage
\renewcommand{\baselinestretch}{1.1} 
% include the next line for double spacing

% \renewcommand{\baselinestretch}{2}

\renewcommand{\epsilon}{\varepsilon}
\def\fixit#1{}
\def\comment#1{}
\def\equno#1{(\ref{#1})}
\def\equnos#1{(#1)}
\def\sectno#1{section~\ref{#1}}
\def\figno#1{Fig.~(\ref{#1})}
\def\D#1#2{{\partial #1 \over \partial #2}}
\def\df#1#2{{\displaystyle{#1 \over #2}}}
\def\tf#1#2{{\textstyle{#1 \over #2}}}
\def\d{{\rm d}}
\def\e{{\rm e}}
\def\i{{\rm i}}
\def\Leff{L_{\rm eff}}

%--------+---------+---------+---------+---------+---------+---------+

Recently some progress has been achieved in extending the gauge theory/supergravity correspondence, originally formulated for superconformal ${\cal N}=4$ SYM theories \cite{malda1, gkp, witt1} to models with fewer supersymmetries and nontrivial RG flow \cite{kn,kt,ks,malda2}. Such models are hoped to be a step in the direction of finding the string duals of realistic gauge theories; indeed, there is reason to believe that the warped deformed conifold background found in \cite{ks}, as well as the similar backgrounds of \cite{malda2,vafa}, capture the correct topology of the string dual to pure ${\cal N}=1$ SYM. The conifold is warped, corresponding to a running of the coupling; deformed, corresponding to chiral symmetry breaking; and the metric in the IR is nonsingular, allowing one to derive the area law for the Wilson loop. Because of these promising features, it is of interest to further investigate such models.

In this paper we concentrate on the warped conifold background and ask the question: can the AdS/CFT correspondence, which gives a concrete prescription for computing gauge theory correlation functions by solving classical SUGRA equations of motion, be extended to this background? The reason there may be a problem is that the metric is no longer even asymptotically AdS, reflecting the nonconformality of the theory. Thus, one might expect to run into difficulties due to UV divergences. Also, one might suspect that at very high energies the theory does not remain a bona fide local 4-dimensional QFT, but flows to something else, like a Little String Theory; this is in fact expected to be true for the background found in \cite{malda2}. Nevertheless, we will see that in our case, at least for the minimal scalar mode corresponding to a gauge theory operator of (bare) dimension 4, we are able to extract a sensible 2 point function following essentially the prescription developed for the conformal case in \cite{gkp, witt1}. The logarithmic corrections to AdS behavior of the geometry will translate to a logarithmic correction to the correlation function, which finds a natural explanation from the gauge theory point of view. It therefore seems plausible that the SUGRA background found in \cite{ks} is the dual of a well-defined, if somewhat unusual, 4-dimensional QFT.

We can think of the gauge theory/SUGRA correspondence in terms of putting a stack of $N$ D3-branes at the singularity of the cone $ds^2 = dr^2 + r^2(dX^5)^2$\, where $X^5$, the base of the cone, is a compact 5-dimensional manifold \cite{kw}. In the D-brane picture, the low-energy theory of this configuration is a SUSY gauge theory with $N$ colors, whose precise gauge group and number of SUSYs depend on the manifold $X^5$. From the SUGRA point of view, the branes warp the surrounding space into a geometry whose metric in the near horizon limit is

\begin{equation}
\label{prodmetric}
ds^2 = h^{-1/2}(r)dx_{\mu}dx^{\mu} + h^{1/2}(r)(dr^2+r^2(dX^5)^2),
\end{equation}
where $x^{\mu}$ are the flat 4-dimensional coordinates, and $h(r) = R^4/r^4$ with $R^4 \sim g_sN(\alpha^{\prime})^2$. The radial coordinate $r$ corresponds to the RG scale with the UV end at large $r$. Note that because of the form of $h$ this is actually a product geometry, $AdS_5 \times X^5$. There is also a Ramond-Ramond self-dual 5-form field strength of the form
\begin{equation}
\label{rr}
F_5 = N(vol(X^5)+*vol(X^5))
\end{equation}
such that $\int_{X^5}F_5=N$. Two well known examples are $X^5 = S^5$ which leads to ${\cal N}=4$ SYM and $X^5 = T^{1,1}$ where the gauge theory is $SU(N) \times SU(N)$ with ${\cal N}=1$ SUSY. 

Let us recall how one extracts gauge theory correlation functions from the dual supergravity background (\ref{prodmetric}). We follow the method of \cite{gkp}. For every SUGRA field $\phi$ there is a corresponding gauge theory operator ${\cal O}$ such that a Schwinger term $W[\phi] = \int d^4x\phi(x){\cal O}(x)$ can be added to the gauge theory action. The gauge theory/SUGRA correspondence then states
\begin{equation}
\label{corr}
\langle e^{-W[\phi(x)]} \rangle = e^{-S[\phi(x)]},
\end{equation}
where $S[\phi(x)]$ is the classical SUGRA action for the field $\phi$ with the boundary condition $\phi(x,\rho)=\phi(x)$ and $\rho$ is an ultraviolet cutoff that can be taken to infinity in the end. We also require that $\phi$ be regular at the IR, i.e. for small $r$. In other words the classical SUGRA action for the field $\phi$ subject to these boundary conditions generates the connected gauge theory correlation functions of the operator ${\cal O}$.

In particular, suppose we want to calculate the two point function $\langle{\cal O}_4(x_1){\cal O}_4(x_2)\rangle$ for an operator ${\cal O}_4$ corresponding to an s-wave (with respect to the compact $X^5$) minimal massless scalar $\phi$ propagating in the geometry (\ref{prodmetric}). The action for such a scalar is
\begin{equation}
\label{action}
S = {1 \over 2\kappa^2}\int d^{10}x{\sqrt g}[{1 \over 2}g^{mn}\partial_m\phi\partial_n\phi] = {V \over 4\kappa^2}\int d^4x\int^{\rho} dr r^5[(\partial_r\phi)^2+h(r)\eta^{\mu \nu}\partial_{\mu}\phi\partial_{\nu}\phi],
\end{equation}
where $V$ is the volume of the $X^5$ and $\rho$ once again is a UV cutoff. The indices $m,n$ run over the entire 10-dimensional space, the indices $\mu,\nu$ over 4-dimensional Euclidean space. The equation of motion resulting from this action is
\begin{equation}
\label{coordeq}
(r^{-5}\partial_rr^5\partial_r+h(r)\eta^{\mu \nu}\partial_{\mu}\partial_{\nu})\phi=0.
\end{equation}
Integrating by parts in the action (\ref{action}), we get
\begin{eqnarray*}
\label{fluxes}
S = {V \over 4\kappa^2}\int d^4x\int^{\rho} drr^5[-\phi(r^{-5}\partial_r r^5\partial_r+h(r)\eta^{\mu \nu}\partial_{\mu}\partial_{\nu})\phi+r^{-5}\partial_r(\phi r^5\partial_r\phi)]= \\
=-{V \over 4\kappa^2}[{\cal F}(r)_{r=\rho}-{\cal F}(r)_{r=0}],
\end{eqnarray*}
where ${\cal F}(r) = \phi(r)r^5\partial_r\phi(r)$ is the flux factor. We have used the equation of motion and the fact that there are no boundary terms from integrating by parts in the $x^{\mu}$ directions since the fields are assumed to vanish at 4-dimensional infinity. Going to momentum space, we find
\begin{equation}
\label{momaction}
S = {V \over 4\kappa^2}\int d^4kd^4q\phi_k\phi_q(2\pi)^4\delta^{(4)}(k+q){\cal F}_k,
\end{equation}
where $\phi(x) = \int d^4k\phi_ke^{ikx}$ and 
\begin{equation}
\label{momflux}
{\cal F}_k = [{\tilde \phi}_kr^5\partial_r{\tilde \phi}_k]_0^{\rho}.
\end{equation}
${\tilde \phi}_k$ are momentum modes normalized to ${\tilde \phi}_k(\rho)=1$. From (\ref{corr}), the corresponding 2 point function in momentum space is then
\begin{equation}
\label{2point}
\langle {\cal O}_4(k){\cal O}_4(q)\rangle = {\partial^2S \over \partial\phi_k\partial\phi_q} = (2\pi)^4\delta^{(4)}(k+q){V \over 4\kappa^2}{\cal F}_k.
\end{equation}
Thus, to extract the 2 point function we need to solve the equations of motion for the momentum $k$ Fourier mode of the field $\phi$ with the boundary conditions $\phi(\rho)=1$, $\phi(r \rightarrow 0)$ regular, and find the flux factor ${\cal F}_k$. Note that we are interested in terms nonanalytic in $k$, since the analytic terms correspond to contact terms in position space.

In the standard AdS/CFT correspondence $h(r)=R^4/r^4$ and the equation of motion (\ref {coordeq}) in momentum space becomes
$$(r^{-5}\partial_rr^5\partial_r-k^2{R^4 \over r^4})\phi =0.$$
Changing variables to $z = R^2/r$ this is
\begin{equation}
\label{besseleq}
(z^3\partial_zz^{-3}\partial_z-k^2)\phi(z)=0.
\end{equation}
This is equivalent to a Bessel equation whose solution with the desired boundary conditions is
$$\phi(z) = {z^2 K_2(kz) \over \varepsilon^2K_2(k\varepsilon)}$$
where $\varepsilon = R^2/\rho$ is a UV cutoff for the $z$ coordinate. This function has the small $z$ expansion
\begin{equation}
\label{besselexpand}
\phi(z) = 1 -{1 \over 4}(kz)^2 - {1 \over 16}(kz)^4\log (kz) + ...
\end{equation}
The logarithmic term gives the leading nonanalytic contribution, so that
$$\langle{\cal O}_4(k){\cal O}_4(-k)\rangle \sim (k\varepsilon)^4 \log (k\varepsilon),$$
or
$$\langle{\cal O}(x_1){\cal O}_4(x_2)\rangle \sim {1 \over |x_1-x_2|^8}.$$

Let us now turn to the warped deformed conifold. First, we take $X^5=T^{1,1}$ which initially gives us ${\cal N}=1$ $SU(N) \times SU(N)$ conformal SYM. $T^{1,1}$ has the topology $S^2 \times S^3$. Now we also wrap $M$ D5-branes over the $S^2$ cycle of the $T^{1,1}$. As shown in \cite{gk,kt,ks}, this results in an ``$SU(N+M) \times SU(N)$'' gauge theory which still has ${\cal N}=1$ SUSY, but is no longer conformal, and undergoes repeated duality cascades. The effective number of colors runs with scale until it is stabilized at a number of order $M$ in the IR. In the IR, the $S^3$ of the $T^{1,1}$ gets blown up to finite size, corresponding to chiral symmetry breaking and confinement. However, if we are interested in high-energy scattering (with respect to the $\Lambda_{QCD}$ scale), we do not need the full details of the nonsingular solution found in \cite{ks}. Instead, we concentrate on the UV part, corresponding to the solution found in \cite{kt}. This solution becomes singular in the IR, but we will impose the boundary condition that the fields do not blow up at this singularity. In effect, we require that the solution we find in the UV be joined smoothly to the IR. This will allow us to extract the 2-point function at sufficiently high energies.

For the warped conifold solution of \cite{kt}, the metric has the form (\ref{prodmetric}) with
\begin{equation}
\label{warpfunc}
h(r) = A^2 {R^4 \over r^4}\log {r \over R},
\end{equation}
where $R$ corresponds to the confinement/chiral symmetry breaking scale, and $A \sim g_sM$. We will not keep track of numerical factors since there is no field theory calculation we can compare our result with. The RR 5-form field strength now takes the form
$$F_5 = {\cal K}(r)vol(T^{1,1})+*{\cal K}(r)vol(T^{1,1})$$
where ${\cal K}(r)=A^2\log(r/R)$, so that the effective number of colors $N$ runs with scale as
\begin{equation}
\label{colors}
N(r) = \int_{T^{1,1}}F_5 = A^2\log {r \over R}.
\end{equation}
Note that the space (\ref{prodmetric}) with $h$ of the form (\ref{warpfunc}) is no longer a product geometry; now there is a logarithmic warping factor. However, because the warping is so mild (logarithmic), we will still be able to extract correlation functions as explained above. We now proceed to do so. 

The equation for the $l=0$ partial wave propagation of a minimal massless scalar in this background is 

\begin{equation}
\label{momeq}
[r^{-5}\partial_r r^5\partial_r -A^2k^2{R^4 \over r^4}\log {r \over R}]\phi(r)=0.
\end{equation}
Changing variables to $y = AkR^2/r$, eq. (\ref{momeq}) becomes
\begin{equation}
\label{yeq}
[y^3\partial_yy^{-3}\partial_y-\log{Y \over y}]\phi(y)=0,
\end{equation}
where $Y = AkR$. The region of interest to us is $y \ll Y$. The extreme UV corresponds to small $y$. For sufficiently small $y$, we solve (\ref{yeq}) by expanding in $y$, and treating the $\log( Y/y)$ term as a perturbation. Namely, we have
$$\phi = \phi_0 +\phi_1 + \phi_2+...$$
where $[y^3\partial_yy^{-3}\partial_y]\phi_{n+1}=[\log(Y/y)]\phi_n$, $\phi_{-1}=0$. As usual, we impose the boundary condition $\phi(0)=1$, where we have already taken the UV cutoff to infinity. We find
\begin{equation}
\label{phiuv}
\phi_{UV}  = 1 - {1 \over 4}y^2\log{Y \over y} + y^4[{1 \over 48}\log^3{Y \over y}+{1 \over 64}\log^2{Y \over y}+{1 \over 128}\log {Y \over y}+C]+...
\end{equation}
where $C$ is an undetermined constant. The information about the 2 point function is hidden in the constant $C$ since all other parts of the above expression are analytic in $k$ (note that $Y/y$ doesn't depend on $k$). To find this constant we follow the equation into the IR. For $y >> 1/Y$, $\log(Y/y) \approx \log Y$. Note that this makes sense since for $k \gg \Lambda_{QCD} \sim 1/(g_sMR)$, $Y$ is a large number. The equation (\ref{yeq}) becomes
\begin{equation}
\label{phiireq}
(y^3\partial_yy^{-3}\partial_y-\log Y)\phi=0.
\end{equation}  
This is Bessel's equation, just like (\ref{besseleq}). Now we impose the condition that $\phi$ is regular in the IR which means that we take the solution that is regular at large $y$. This is the same solution as we needed in (\ref{besselexpand}):
\begin{equation}
\label{phiir}
\phi_{IR} = D(1 - {1 \over 4}y^2\log Y-{1 \over 16}y^4\log^2 Y\log(\sqrt {\log Y}y)+...)
\end{equation}
We will now match $\phi_{UV}$ to $\phi_{IR}$. Let us first identify the overlap region. We said before that the solution (\ref{phiuv}) is valid for small $y$. By looking at this solution we see that it has the form of an expansion in $y^2\log(Y/y)$, so we are allowed to use this solution when $y^2\log(Y/y) \ll 1$. On the other hand, the condition for the validity of eq. (\ref{phiireq}) is $y \gg 1/Y$. We see that when $Y$ is large, these conditions are compatible and there is an overlap region $1/Y \ll y \ll 1/\sqrt{\log Y}$. In this region we can drop the $\log y$ terms in (\ref{phiuv}) since $|\log y| \ll |\log Y|$. Matching (\ref{phiuv}) to (\ref{phiir}) order by order, the first two terms match if we set $D=1$. However, if we look at the terms multiplying $y^4$, we see that $\phi_{UV}$ has a $\log^3 Y$ term, whereas the leading term in $\phi_{IR}$ is a $\log^2 Y\log\log Y$ term. Luckily, we have an undetermined constant $C$ that we can use to cancel this leading $\log^3 Y$ term. Thus, we find
\begin{equation}
\label{const}
C = -{1 \over 16}\log^3 Y + ...=-{1 \over 16}\log^3 AkR + ...
\end{equation}
where we have kept only the leading nonanalytic term. Using equations (\ref{phiuv},\ref{const},\ref{momflux},\ref{2point}), we are now ready to compute the 2 point function. It is
\begin{equation}
\label{mom2point}
\langle {\cal O}_4(k){\cal O}_4(-k)\rangle \sim {A^4k^4 \over \kappa^2} \log^3 AkR.
\end{equation} 
Here $\kappa$ is the 10-dimesntional gravitational coupling. This translates into a position space 2 point function that behaves like
\begin{equation}
\label{pos2point}
\langle{\cal O}_4(x_1){\cal O}_4(x_2)\rangle \sim g_s^2M^4{\log^2(|x_1-x_2|^2/(g_sMR)^2) \over |x_1-x_2|^8}.
\end{equation}
This is our result, which is valid in the UV range ($|x_1-x_2| \ll g_sMR$). It seems to indicate that one should be able at least in principle to extend the standard methods of the AdS/CFT correspondence to this nonconformal model and extract sensible correlation functions. Note that for large $g_sM$, that is, when the SUGRA description is valid, this gauge theory is at strong coupling {\it for all scales}\cite{ks}. Thus, there is no perturbative field theory calculation that we can compare (\ref{pos2point}) to. However, if we look at the way the effective number of colors in the theory runs with scale (\ref{colors}), we notice that (\ref{pos2point}) can be written as
$$\langle{\cal O}_4(x_1){\cal O}_4(x_2)\rangle \sim {N_{eff}^2(|x_1-x_2|) \over |x_1-x_2|^8}.$$
Since we expect all correlation function to be multiplied by a factor of $N^2$, this makes sense. We see that the logarithmic correction we find to conformal behavior is explained by the logarithmic running of the effective number of colors. So it seems that the dual to this SUGRA background remains a local 4-dimensional quantum field theory at arbitrarily high energies.

\begin{figure}
\epsfbox{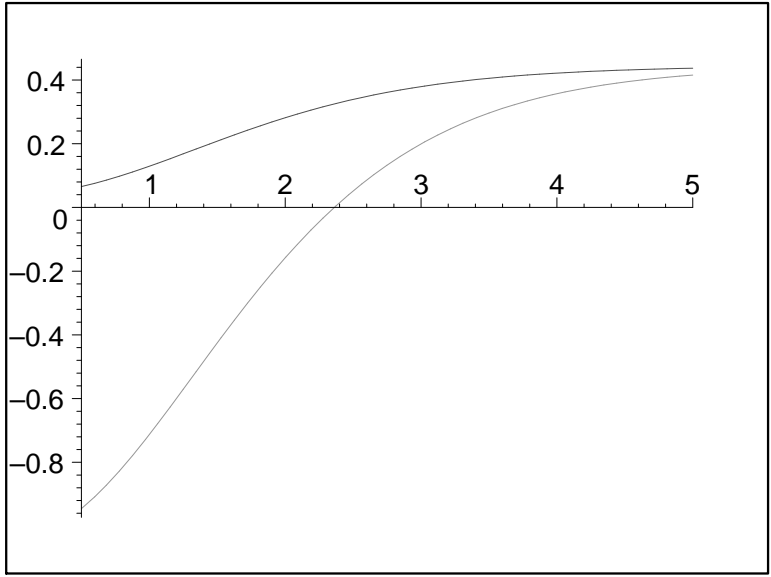}
\caption{${\tilde k}^2(\tau)$ vs. $\tau$ for $m^2=0.5$ (solid line) and $m^2=2$ (dashed line)}
\label{gluefig}
\end{figure}

As we already emphasized, the form (\ref{pos2point}) of the 2 point function is valid only in the UV. Its large distance behavior is obtained by considering the low energy solutions; we will see that these exhibit a mass gap, so at large distances the 2 point function falls off exponentially. To demonstrate the existence of a mass gap, we compute the spectrum of low-lying glueball modes. This calculation proceeds as follows\cite{witt2,coot,mjmn}: we need to go from Euclidean to Minkowski space, and find solutions to eq. (\ref{coordeq}) that are normalizable both in the UV and the IR. We will see that such solutions exist only for a discrete set of $k^2$ which give the glueball masses. The IR can no longer be ignored, so we need to use the full background of \cite{ks}. The full metric is now given by
\begin{equation}
\label{full}
ds^2 = h^{-1/2}(\tau)dx_{\mu}dx^{\mu}+h^{1/2}(\tau)ds_6^2,
\end{equation}
where the metric of the deformed conifold is given by\cite{ohta}
\begin{equation}
\label{6form}
ds_6^2 = {3{\tilde R}^2 \over 2^{2/3}}K(\tau)[{1 \over 3 K^3(\tau)}(d\tau^2+(g^5)^2)+\cosh^2(\tau/2)((g^3)^2+(g^4)^2)+\sinh^2(\tau/2)((g^1)^2+(g^2)^2)].
\end{equation}
The $g^i$ are forms depending only on the angular variables on $T^{1,1}$ and the functions $h$ and $K$ are given by
\begin{eqnarray}
\label{hkform}
h(\tau) = {\tilde A}^2\int_{\tau}^{\infty}dx{x\coth x-1 \over \sinh^2x}(\sinh(2x)-2x)^{1/3}, \\
K(\tau) = {(\sinh(2\tau)-2\tau)^{1/3} \over 2^{1/3}\sinh\tau}.
\end{eqnarray}
$\tau$ is a dimensionless radial variable, which in the UV is related to our previous variable $r$ by $r \sim Re^{-\tau/3}$. Also, ${\tilde A}\sim g_sM$ and ${\tilde R}$ are related to $A$ and $R$ by numerical factors. The equation of motion for a massless minimal scalar in the metric (\ref{full}) is 
\begin{equation}
\label{glueeq} [(\sinh 2\tau-2\tau)^{-2/3}\partial_{\tau}(\sinh 2\tau-2\tau)^{2/3}\partial_{\tau}+{(m{\tilde R})^2\sinh^2\tau \over (\sinh 2\tau-2\tau)^{2/3}}h(\tau)]\phi=0.
\end{equation}
Here $m^2=k^2$ is the mass squared of the 4-dimensional mode. Our goal is to find values of $m^2$ for which eq. (\ref{glueeq}) has solutions that are normalizable at both the UV and the IR. This means that the flux factor
\begin{equation}
\label{glueflux}
{\cal F} = \phi(\tau)(\sinh2\tau-2\tau)^{2/3}\partial_{\tau}\phi(\tau)
\end{equation}
must remain finite at both $\tau=\infty$ and $\tau=0$ (in fact for the normalizable solutions it will vanish at both ends). Let $f(\tau) = (\sinh2\tau-2\tau)^{2/3}$, and define the new field $\psi(\tau)$ by $\phi(\tau)=f^{-1/2}(\tau)\psi(\tau)$. Then the equation (\ref{glueeq}), written in terms of $\psi$, reduces to the more familiar ``Schroedinger'' form
\begin{equation}
\label{schroed}
[\partial_\tau^2-{\tilde k}^2(\tau)]\psi = 0,
\end{equation}
where
\begin{equation}
\label{ktilde}
{\tilde k}^2(\tau) = {4 \over 3}{\sinh 2\tau \over \sinh 2\tau - 2\tau}-{8 \over 9}{(\cosh 2\tau -1)^2 \over (\sinh 2\tau -2\tau)^2}-{(m{\tilde R})^2\sinh^2\tau \over (\sinh 2\tau-2\tau)^{2/3}}h(\tau)
\end{equation}
with the flux (\ref{glueflux}) now expressed as
\begin{equation}
\label{schroedflux}
{\cal F} = \psi\partial_{\tau}\psi-{1 \over 2}(\partial_{\tau}\log f)\psi^2.
\end{equation}
At $\tau \rightarrow \infty$, ${\tilde k}^2 \rightarrow 4/9$ and we have the solutions $\psi_{\pm} \sim e^{\pm 2\tau /3}$; only $\psi_-$ is normalizable. At $\tau \rightarrow 0$, the normalizable solution behaves as $\psi \sim \sinh({\tilde k}(0)\tau)$, where
\begin{equation}
\label{k0}
{\tilde k}^2(0) = {2 \over 5} - m^2({\tilde A}{\tilde R})^2\int_0^{\infty}dx{x\coth x-1 \over \sinh^2x}(\sinh(2x)-2x)^{1/3}.
\end{equation}
We now find the eigenvalues of eq. (\ref{schroed}) using the WKB approximation (see e.g. \cite{schiff}), which gives a sensible estimate for a smooth potential like ${\tilde k}^2$ and is increasingly accurate for more excited states. In this approximation, (\ref{schroed}) has the solutions
$$\psi_{\pm}(\tau) \sim {\tilde k}^{-1/2}(\tau)e^{\pm \int^{\tau}{\tilde k}(x)dx}$$
which are valid away from the turning points ${\tilde k}=0$. At a turning point $\tau_0$, the exponentially decreasing solution on one side is matched to an oscillatory solution on the other through
$${\tilde k}^{-1/2}e^{-\int^{\tau}{\tilde k}(x)dx} \rightarrow k^{-1/2}\cos (\zeta(\tau)-\pi /4),$$
where $k^2 = -{\tilde k}^2$ when ${\tilde k}^2 < 0$, and
$$\zeta(\tau) = \int_{\tau}^{\tau_0} k(x)dx.$$
In our case, it's not hard to see that the function ${\tilde k}^2(\tau)$ increases monotonically with $\tau$ from its value ${\tilde k}^2(0) < 2/5$ to ${\tilde k}^2(\infty)=4/9$ (see fig. \ref{gluefig}).
When $m^2$ is small enough, ${\tilde k}^2(0) > 0$ and there are no turning points at all. Then the exponentially decreasing solution, which is not normalizable at the IR, is valid for all $\tau$, so there are no normalizable modes; in other words, there is a mass gap, as advertised. For sufficiently large $m^2$ there are normalizable modes determined by the transcendental equation
\begin{equation} 
\label{trans}
\int_0^{\tau_0(m^2)} k(x)dx = {3 \pi \over 4} +(n-1)\pi
\end{equation}
where $n=1,2,3...$ and $\tau_0$ is given by ${\tilde k}^2(\tau_0)=0$. The phase in (\ref{trans}) must be such that $\psi$ behaves as a pure sine wave near $\tau = 0$. Solving this equation numerically we obtain the glueball modes:

$$m_n^2 = c_n ({\tilde A}{\tilde R})^{-2}$$
with the first few coefficients being
$$c_1 = 1.79, c_2 = 4.03, c_3 = 7.16, c_4 = 11.2, c_5 = 16.2, c_6 = 22.0, c_7 = 28.8$$
and so on. The overall normalization of these coefficients is of course arbitrary but their ratios are dimensionless numbers that constitute a prediction for the gauge theory. As expected, we have $m^2 \sim \Lambda_s^2/(g_sM)^2$, where $\Lambda_s \sim 1/R$ is the string tension scale. This is dimensional transmutation.

There is more information about this cascading gauge theory to be extracted from its SUGRA dual. In a recent paper \cite{buchel}, a finite temperature solution was obtained corresponding to a black hole in the background of \cite{kt}. The entropy of the black hole was found to scale with the temperature as $S \sim N^2(T) T^3 \sim  (\log^2 T)T^3$ which is consistent with our picture of the number of colors $N$ running with the scale $\mu$ as $N \sim \log \mu$. Other possibilities include identifying further (perhaps nonminimal) modes in the SUGRA background and computing the corresponding correlators. Such calculations might allow a more delicate probe of the RG structure of this theory. We leave them for future work.

\section*{Acknowledgments}

I am grateful to Igor Klebanov for suggesting this problem and for many helpful discussions. Thanks also to Chris Herzog and Leonardo Rastelli for commenting on drafts of this paper.

\end{document}